\documentclass[twocolumn,showpacs,preprintnumbers,nofootinbib]{revtex4}

\usepackage{epsf}
\usepackage{graphicx}
\usepackage{dcolumn}
\usepackage{bm}
\def\be{\begin{equation}}
\def\ee{\end{equation}}
\def\bea{\begin{eqnarray}}
\def\eea{\end{eqnarray}}

\newlength{\dinwidth}
\newlength{\dinmargin}
\def\fun#1#2{\lower3.6pt\vbox{\baselineskip0pt\lineskip.9pt
  \ialign{$\mathsurround=0pt#1\hfil##\hfil$\crcr#2\crcr\sim\crcr}}}
\begin{document}
 \tighten
\vskip 3cm

\

\title{Creation of a Compact Topologically Nontrivial Inflationary Universe}
\author{ Andrei Linde}
\address{Department
  of Physics, Stanford University, Stanford, CA 94305-4060,
USA    }

\date{September 10, 2004}
{\begin{abstract} If inflation can occur only at the energy density V much
smaller than the Planck density, which is the case for many inflationary models
based on string theory, then the probability of quantum creation of a closed or
an infinitely large open inflationary universe is exponentially suppressed for
all known choices of the wave function of the universe. Meanwhile under certain
conditions there is no exponential suppression for creation of topologically
nontrivial compact flat or open inflationary universes. This suggests, contrary
to the standard textbook lore, that compact flat or open universes with
nontrivial topology should be considered a rule rather than an exception.
\end{abstract}}

 \maketitle

\section{Introduction}

The standard textbook description of the cosmological models usually is limited
to the description of the three Friedmann models, describing a closed universe,
and infinite topologically trivial flat and open universes. However, in his
papers Friedmann noted that even flat and open universes may be compact if they
have nontrivial topology \cite{Friedmann}. He described how space could be
finite and multi-connected by suitably identifying  points, and predicted the
possible existence of ``ghost" images of astronomical sources. The possibility
that we may live in a compact universe with nontrivial topology was mentioned
even earlier  by Schwarzschild \cite{Schw1900}. Since that time, there were
dozens of papers on the compact flat or open universes, see e.g.
\cite{ZelStar,gb,chaotmix,topol4,topol1,topol2,Barrow,topol3}. Nevertheless,
this possibility is not mentioned even as a footnote in most of the textbooks
on cosmology and general theory or relativity.

This attitude is quite understandable. First of all, we still do not have a
compelling observational evidence of the nontrivial topology of the universe
\cite{Cornish:2003db}, even though the situation may change with the further
investigation of the CMB anisotropy. Moreover, inflation typically makes the
universe so large that it removes any hope for the observational study of its
topological structure, unless one fine-tunes the total number of e-folds of
inflation and bends the overall shape of the scalar potential in a rather
peculiar way \cite{Linde:2003hc}. So one may wonder why should we spend our
time studying topology of the universe if it is not supported by experimental
evidence and theoretical expectations?

As we will argue in this paper, the models of a compact flat or open
(hyperbolic) universe with nontrivial topology may play an important role in
inflationary cosmology by providing proper initial conditions for low-scale
inflation. Most of our arguments will be based on results obtained in the
papers on quantum creation of the  compact universe with nontrivial topology
\cite{ZelStar,gb} and in the papers on chaotic mixing \cite{chaotmix,topol4}.
However, we will look at these results from a slightly different perspective:
We will try to find out how the probability of initial conditions for inflation
depends on the effective potential of the inflaton field $V(\phi)$ for various
versions of topology of the universe.

\section{The problem of initial conditions for the low-scale
inflation}\label{problem}

Let us first remember some basic facts related to the
problem of initial conditions for inflation \cite{book}. Inflation appears in a
patch of a closed universe or of an infinite flat or open universe if this
patch is sufficiently homogeneous and its size is greater than the size of the
horizon, $\Delta l \gtrsim H^{-1} \sim V^{-1/2}(\phi)/M_p$, where $V(\phi)$ is
a flat inflationary potential.

The simplest inflationary scenario is based on the theory of a scalar field
with the potential $\sim \phi^n$. Formally, inflation in this scenario can
start at an arbitrarily large values of $\phi$, i.e. close to the cosmological
singularity. In this paper we will use the system of units $M_p^2 =(8\pi
G)^{-1} = 1$. If we assume that the standard description of the universe in
terms of the classical space-time becomes possible only below the Planck
density $M_p^4 = 1$, then inflation may begin, for example, in a closed
universe of the smallest possible size $\Delta l =O(1)$ with the total entropy
$S =O(1)$. These conditions seem quite natural and easy to satisfy; see e.g. a
discussion of this issue in \cite{book}.

However, in all inflationary models where the effective potential does not
change too much during the last stages of inflation one has a general
constraint $V \lesssim 10^{-8}$, which follows from the constraint on the
amplitude of gravitational waves generated during inflation. In many
inflationary models, such as new inflation, hybrid inflation, etc., inflation
occurs at $V(\phi) \ll 10^{-8}$. In these models, the problem of initial
conditions becomes more complicated. Consider, for example, a closed universe
filled with radiation. At the Planck time, the energy density of radiation $T^4
\sim 1$ was completely dominated the total energy density $\sim T^4 +V(\phi)$.
Inflation begins only when the universe cools down and the energy density of
radiation becomes much smaller than $V(\phi)$, which happens at $t \sim H^{-1}
\sim V^{-1/2}$. The problem is that a radiation dominated closed universe may
die before this happens.

Indeed, a closed universe with energy density dominated by radiation collapses
within the time $t \sim S^{2/3}$ (in Planckian units), where $S$ is the total
entropy of the universe \cite{book}. If we want the universe to survive until
its energy density drops down to $V(\phi)$, its initial entropy should be
greater than $S \sim V^{-3/4}$. This means, e.g., that for the models where
inflation may occur only at $V< 10^{-12}$, the initial entropy of the universe
must be $S> 10^9$. At the Planck time the total mass of such universe would be
greater than $10^9$ in Planck units, and it would consist of $10^9$ causally
disconnected regions of Planck size. Thus in order to explain why our universe
is so large and homogeneous we would need to assume that it was very large and
homogeneous from the very beginning \cite{book}.  An estimate of the
probability of such event for the universe initially dominated by relativistic
matter suggests that it should be suppressed by a factor of $\exp(-S)\sim
\exp(-V^{-3/4}) < e^{-10^9}$, and the probability that not only the energy
density but also the scalar field in such a universe will be sufficiently
homogeneous may be even smaller   \cite{Linde:1994wt}.

Another way to look at the same problem is to assume that instead of being born
in a singularity and passing through the radiation dominated stage, a closed
universe was created ``from nothing'' (i.e. from a state with a scale factor $a
= 0$) in an inflationary state dominated by the inflaton potential energy
$V(\phi)$. However, the minimal size of a closed inflationary universe is given
by $H^{-1} \sim V^{-1/2}$. Therefore the universe with $V(\phi)\ll 1$ cannot
classically evolve starting from the point $a=0$. Since this process is
forbidden at the classical level, the universe must tunnel through a potential
barrier. An investigation of this issue using Euclidean approach to quantum
cosmology  suggested \cite{Linde:1983mx,Vilenkin:1984wp} that the probability
of quantum creation of a closed inflationary universe is exponentially
suppressed by a factor of
\begin{equation}\label{1}
P \sim \exp\Bigl(-{24\pi^2\over V(\phi)}\Bigr) \ .
\end{equation}

This result has an obvious interpretation. Creation of the universe from
nothing implies emergence of a universe with initial volume $H^{-3}$ and with
total energy of matter $\Delta E \sim H^{-3}V \sim V^{-1/2}$. For $V \sim 1$
(Planck density), uncertainty relations tell us that such fluctuations can
readily occur within the Planck time $\Delta t = O(1)$, but for $V\ll 1$ these
fluctuations should be extremely improbable, which is reflected in Eq.
(\ref{1}). In a particular example with $V \sim 10^{-12}$, the probability of
quantum creation of inflationary universe is suppressed\footnote{The
Hartle-Hawking wave function \cite{Hartle:1983ai} yields a different result, $P
\sim \exp(+{24\pi^2\over V(\phi)})$. In this case, the probability of inflation
is also exponentially suppressed, as compared with the probability of living in
the deepest minimum of $V(\phi)$. In our opinion,  this function, as well as
its recent modifications \cite{Firouzjahi:2004mx}, describe the probability of
the {\it final} conditions in the universe, instead of the {\it initial}
conditions; see a discussion of this issue in \cite{book,Linde:1998gs}. It is
important that in either interpretation, the exponential suppression of
probability is due to the tunneling, i.e. due to the classically forbidden
evolution. } by the factor of $10^{-10^{14}}$.

Open and flat inflationary universes can classically evolve from $a = 0$
without any tunneling. However, we do not know any way to create an infinite
homogeneous flat universe. There are two different ways to create an infinite
homogeneous open universe. The first one involves creating a closed
inflationary universe of the old inflation type. Then one must ensure that it
decays by bubble formation and that the slow-roll  inflation occurs inside each
bubble \cite{Linde:1995rv}. Another possibility is to create an infinite open
universe directly, using a different analytic continuation of the instanton of
the same type as  the one used for the description of creation of a closed
universe \cite{Hawking:1998bn}. Both cases involve tunneling and exponential
suppression of the probability of creation of an inflationary universe with
$V\ll 1$ \cite{Linde:1998gs}.

\section{Low-scale inflation in a compact open or flat universe}
In the previous section we identified the root of the problem of initial
conditions for creation of a closed or an infinitely large open inflationary
universe: If the universe initially was matter dominated, it should contain a
very large nearly homogeneous patch with large initial mass and entropy. If, on
the other hand, we consider a possibility of a quantum creation of the
universe, the probability of such an event is exponentially suppressed because
a  part of the trajectory describing creation of the universe is forbidden at
the classical level.

However, in a compact flat or open universe none of these problems exist, under
certain conditions to be discussed below.

\subsection{Creation of a hot compact universe}
Consider for simplicity the flat compact universe having the topology of a
torus, $S_1^3$,
\begin{equation}\label{2}
ds^2 = dt^2 -a_i^2(t)\,dx_i^2
\end{equation}
with identification $x_i+1 = x_i$ for each of the three dimensions. Consider
for  simplicity the case $a_1 = a_2 = a_3 = a(t)$. In this case the curvature
of the universe and the Einstein equations written in terms of $a(t)$ will be
the same as in the infinite flat Friedmann universe with metric $ds^2 = dt^2
-a^2(t)\,d{\bf x^2}$. In our notation, the scale factor $a(t)$ is equal to the
size of the universe in Planck units $M_{p}^{{-1}} = 1$.

Let us assume, as we did in the beginning of the previous section, that at the
Planck time $t_p \sim M_p^{-1}=1$ the universe was radiation dominated, $V\ll
 T^4 = O(1)$. Let us also assume that at the Planck time the total size of the
box was Planckian, $a(t_p) = O(1)$. In such case the whole universe initially
contained only $O(1)$ relativistic particles such as photons or gravitons, so
that the total entropy of the whole universe was O(1).

In general, in addition to the energy of relativistic particles one should also
consider the Casimir energy which appears due to the finiteness of the size of
the box. However, Casimir energy is suppressed by supersymmetry. In the absence
of supersymmetry,  this energy is $O(a^{-4})$, so if $a$ is somewhat greater
than $1$ at the Planck density, this energy always remains smaller than the
thermal energy.

The size of the universe dominated by relativistic particles was growing as
$a(t) \sim \sqrt t$, whereas the mean free path of the gravitons was growing as
$H^{-1}\sim t$. If the initial size of the universe was $O(1)$, then at the
time  $t \gg 1$ each particle (or a gravitational perturbation of metric)
within one cosmological time would run all over the torus many times, appearing
in all of its parts with nearly equal probability. This effect, called
``chaotic mixing,'' should lead to a rapid homogenization of the universe
\cite{chaotmix,topol4}. Note, that to achieve a modest degree of homogeneity
required for inflation to start when the density of ordinary matter drops down,
we do not even need chaotic mixing. Indeed, density perturbations do not grow
in a universe dominated by ultrarelativistic particles if  the size of the
universe is smaller than $H^{-1}$. As we just mentioned, this is exactly what
happens in our model.

Note that this homogenization applies to all light fields, including  the
inflaton field. Consider, for example, an initial domain of a Planckian size
$O(1)$, containing the inhomogeneities of the scalar field with the energy
density $(\partial_{i}\phi)^{2}\sim T^{4} \leq 1$. Since the mass squared of
the inflaton field is much smaller than $V$, and the size of the universe prior
to inflation is smaller than $H^{-1}$,  perturbations of this field behave as
ultrarelativistic particles, the distribution of density of scalar
perturbations becomes nearly homogeneous, and the energy density of the
gradients of the scalar field will decrease as the energy of radiation (as
$a^{-4}$) all the way until the beginning of inflation when $V(\phi)$ begins to
dominate.

Chaotic mixing can be very efficient in a compact hyperbolic universe (i.e. in
a compactified version of an open universe) \cite{chaotmix,topol4}.  Moreover,
the effective energy associated with the curvature of 3D space in the open
universe case decreases as $a^{-2}$, i.e. much more slowly than the energy of
ordinary matter. Soon after the beginning of expansion, the curvature energy
density becomes dominant,  and the relative role of normal matter and of its
density perturbations become insignificant. This is an additional reason to
expect that the density perturbations in a compact hyperbolic universe will
rapidly decrease \cite{Mukh}.  Therefore the universe should remain relatively
homogeneous until the thermal energy density drops below $V$ and inflation
begins.

One may wonder whether we need inflation at all if a compact universe can be
efficiently homogenized. But the main reason why it becomes homogeneous is
because we were able to start with a universe of a very small size. In a
non-inflationary universe of the present size, chaotic mixing would not have
enough time to occur since the present part of the universe only now becomes
causally connected. If one starts with the universe with a typical 3D curvature
$O(1)$, then in the absence of inflation our universe would have vanishingly
small $\Omega$. Finally, in the absence of inflation one would have no
explanation to the large entropy of the observable part of the universe, $S >
10^{87}$, and we would have problems explaining the origin of the large scale
structure of the universe, as well as the CMB anisotropy.

Thus without the help of inflation the effects considered above cannot solve
the major cosmological problems. However, they can keep the universe
sufficiently homogeneous until the onset of inflation.

The authors of Ref. \cite{topol4} discussed conditions which could lead to
damping of initial density perturbations down to $10^{-5}$, with the goal to
have only a relatively short stage of inflation and obtain a model of a
homogeneous universe with $\Omega \sim 0.3$. Our goals are much more modest: We
only need the initial expanding domain of the universe to be sufficiently
homogeneous (${\delta\rho\over \rho} \lesssim 1$) for the onset of the
subsequent stage of inflation. It seems rather easy to achieve such a damping
of density perturbations at the radiation dominated stage in a domain of the
horizon size, or of the size smaller than $H^{-1}$. One could not use this
mechanism in the non-inflationary models of a closed or open universe because
our universe previously consisted of many causally disconnected domains of
horizon size, but one can do it during the pre-inflationary stage in our
scenario.

Similarly, it was shown long ago that particle production in the very early
universe can make the universe locally isotropic \cite{Zeldovich:1971mw}.
However, this process could not make the universe globally isotropic because in
the hot big bang scenario the universe consisted of many causally disconnected
domains. But such effects may play important role in our scenario by making the
initial small patch of the universe isotropic, and then this isotropy will be
further enhanced by the subsequent stage of inflation.

It is interesting to estimate the size of a flat universe of initial Planckian
size at the moment when inflation begins. This happens when $T^4 \sim a^{-4}$
drops from $O(1)$ to $V(\phi)$, i.e. at $a \sim V^{-1/4}$. Note that in this
case the size of the universe at the beginning of inflation is much smaller
than the size of dS horizon $H^{-1}\sim V^{-1/2}$. This is a counterexample to
the common lore that the size of inflationary universe must be greater than
$H^{-1}$.

The probability of creation of a small part of the universe of a size $O(1)$ at
the Planck density is not expected to be exponentially suppressed. Most
importantly, since at that time the evolution of the universe is dominated by
the relativistic matter, this probability practically does not depend on
$V(\phi)$. After that, inflation occurs in an expanding and cooling universe
automatically, if the inflaton field did not strongly interact with matter and
if from the very beginning it was staying at the flat (inflationary) region of
the effective potential. Of course, there will be some suppression of the
probability of inflationary initial conditions if an inflationary regime occurs
only for a narrow range of initial values of the inflaton field. However, this
phase space suppression is much milder than the exponential suppression of the
type described by Eq. (\ref{1}). One can argue that this suppression can be
easily compensated by the exponential growth of volume of the inflationary
domains.

We conclude that in this class of cosmological models, unlike in the usual case
of a closed or an infinite open universe, {\it the probability of inflation is
not exponentially suppressed for the theories with $V(\phi)\ll 1$}.

\subsection{Quantum creation of a compact inflationary universe}

Now we will study a possibility that the universe from the very beginning is
dominated by the potential energy density of the scalar field. As we already
discussed, in the closed universe case such solutions require tunneling from $a
=0$ to $a = H^{-1} = \sqrt {3\over V}$. One way to see it is to study the
well-known Wheeler-DeWitt equation for a closed de Sitter (dS) universe
\cite{DeWitt67,Hartle:1983ai,book}:
\begin{equation}\label{4}
\left[{1\over 24\pi^2}{d^2\over da^2} -6\pi^2 a^2 + 2\pi^2 a^4 V\right]\Psi(a)
= 0
\end{equation}
In this equation we temporarily ignored the evolution of the scalar fields (we
will discuss it later, see Eqs. (\ref{complete}), (\ref{5a})), and assumed that
$V(\phi) = {\rm const}$, which is a good approximation in inflationary
cosmology if the potential $V$ is sufficiently flat. This equation describes
tunneling from $a = 0$  to point $a = \sqrt {3\over V}$ through the potential
barrier ${\cal V}(a) =6\pi^2 a^2 - 2\pi^2 a^4 V$. Meanwhile for a flat dS
universe with a toroidal compactification the cosmological evolution can begin
at an arbitrarily small $a$, i.e. there is no need to tunnel through a
classically forbidden region of $a$ \cite{ZelStar}. As pointed out by Zeldovich
and Starobinsky, in this case, unlike in the case of the topologically trivial
flat dS universe, the flat compactified dS is geodesically complete
\cite{ZelStar}.

In order to derive the Wheeler-DeWitt equation for the compact flat toroidal
universe, one should first consider the gravitational action $S = \int dt\, d^3
x\,\sqrt{-g}\,(-{1\over 2}{R} +{1\over
2}\partial_{\mu}\phi\,\partial^{\mu}\phi-V(\phi))$ and take into account that
the volume of the 3D box is equal to $a^{3}$. (In a closed universe the volume
is given by $2\pi^{2}a^{3}$. That is why the coefficients in the Wheeler-DeWitt
equation for the toroidal universe will be different from the analogous
coefficients for the closed universe.) Let us assume for a moment that $\phi$
is constant (see below). In this case one can represent the effective
Lagrangian for the scale factor as a function of $a$ and $\dot a$,
\begin{equation}\label{lagr}
L(a) = -3\dot a^{2} a - a^{3}V\ .
\end{equation}
Finding the corresponding Hamiltonian and using the Hamiltonian constraint $H
\Psi(a) =0$ yields the Wheeler-DeWitt equation
\begin{equation}\label{5}
\left[{d^2\over da^2} +12 a^4 V\right]\Psi(a) = 0
\end{equation}
Eqs. (\ref{5}) differs from Eq. (\ref{4}) not only by numerical coefficients,
but mainly by the absence of the term $6\pi^2 a^2$, which was leading to the
existence of the barrier in the effective potential ${\cal V}(a)$ for the
closed universe. In the flat toroidal universe case there is no such barrier.

One should note that here we ignored the Casimir effect, which may give an
important quantum contribution $O(a^{-4})$ to the vacuum energy density
\cite{ZelStar,gb}. Under certain conditions, this contribution may be negative,
which will again forbid the classical evolution at $a < O(H^{-1})$. However,
this effect is suppressed by supersymmetry \cite{ZelStar,gb}, it disappears in
some anisotropic versions of toroidal compactification  \cite{ZelStar}, and, as
we mentioned in the previous section, it becomes unimportant if one takes into
account the usual matter contribution, or the 3D curvature in the open universe
case\footnote{During the last 20 years the idea of necessity of tunneling at
the moment of the universe creation was so popular that the main effort of the
investigation of Casimir effect in supergravity in \cite{gb} was devoted to
finding versions of supergravity with twisted fields, which break supersymmetry
and lead to the negative Casimir energy and to the exponential suppression of
the tunneling.}.

For large $a$, the solution of Eq. (\ref{5}) can be easily obtained in the WKB
(semiclassical) approximation, $\Psi \sim a^{{-1}}\exp [\pm i {2a^3\sqrt
V\over\sqrt 3}]$; positive sign corresponds to an expanding universe. This
approximation breaks down at $a \lesssim V^{-1/6}$ \cite{ZelStar}. At that time
the size of the universe is much greater than the Planck scale, but much
smaller than the Hubble scale $H^{-1}\sim V^{-1/2}$. The meaning of this
result, to be discussed below in a more detailed way, is that at $a \gg
V^{-1/6}$ the effective action corresponding to the expanding universe is very
large, and the universe can be described in terms of classical space and time.
Meanwhile at $a \lesssim V^{-1/6}$, the effective action becomes small, the
classical description breaks down, and quantum uncertainty becomes large. In
other words, contrary to the usual expectations, at $a \lesssim V^{-1/6}$ one
cannot describe the universe in terms of a classical space-time even though the
size of the universe at $a \sim V^{-1/6}$ is much greater than the Planck size,
and the density of matter as well as the curvature scalar in this regime
remains small, $R = 4 V \ll 1$.

\begin{figure}[h!]
\vskip 0.1cm \centerline{\leavevmode\epsfysize= 5 cm \epsfbox{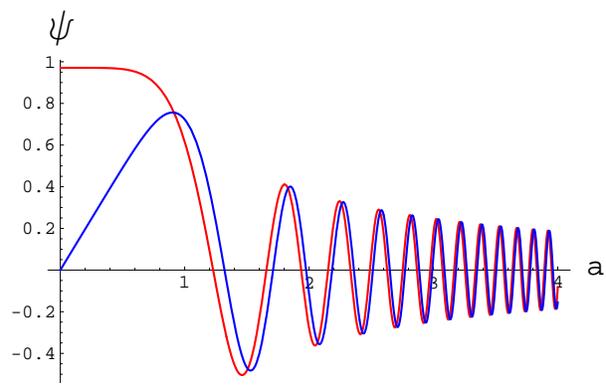}}
\caption[1]{\label{wdw} Two eigenmodes of the Wheeler-DeWitt equation for the
wave function of the flat compact toroidal universe. The scale factor $a$ is
given in units of $V^{-1/6}$. The description of the evolution of the universe
in  terms of classical space-time begins at $a\gtrsim V^{-1/6}$ ($a \gtrsim 1$
in our figure), when the semiclassical approximation becomes valid and the
``cosmic clock'' starts ticking. }
\end{figure}

In this regime one should go beyond the WKB approximation. The general solution
for Eq. (\ref{5}) can be represented as a sum of two Bessel functions:
\begin{eqnarray}\label{funct} \Psi(a) =\beta\sqrt a \Bigl({\rm J}_{-{1\over
6}}\Bigl({2\sqrt V a^3\over \sqrt 3}\Bigr) + \gamma \, {\rm J}_{{1\over
6}}\Bigl({2\sqrt V a^3\over \sqrt 3}\Bigr)\Bigr),
 \end{eqnarray}
where $\beta$ and $\gamma$ are some complex constants. These functions are
shown in Fig. \ref{wdw}.

The expression for the probability to have a universe with a scale factor $a$
is given by $j(a) = -{i\over 2} (\Psi^*\partial_a\Psi- \Psi\partial_a\Psi^*)$
\cite{DeWitt67,Vil89}. An example of the function $\Psi$ describing an
expanding universe and satisfying the normalization condition $j(a) =1$ is
given by Eq. (\ref{funct}) with $\beta = \sqrt{2\,\Gamma(5/6)\, \Gamma(7/8)}$
and $\gamma = -e^{-i\pi/ 6}$.

As we see from this figure, quantum mechanical description of the universe near
$a = 0$ is quite regular. The universe appears ``from nothing'' without
tunneling and without exponential suppression of the probability of its
creation. A similar result is valid for the open universe as
well.\footnote{After this work was completed we became aware of Ref.
\cite{Coule}, where it was also found that the probability of quantum creation
of compact flat and open dS universes (in the case $K =0$, see below) is not
exponentially suppressed, which agrees with \cite{ZelStar} and with our
results. }

One can provide an alternative interpretation of this result, without invoking
the Wheeler-DeWitt equation.  By substituting the classical solution $a =
e^{Ht}$ into the effective Lagrangian (\ref{lagr}), one finds the total action
of the universe:\begin{equation}\label{action} S(t) = -{2\sqrt V\over\sqrt{3}}
a^{3}(t) =  -{2\sqrt V\over\sqrt{3}} e^{{3Ht}}.
\end{equation}
As we see, for $a < V^{-1/6}$ the action becomes smaller than $1$, which means
that one must describe evolution of the universe quantum mechanically. This is
what we did above. However, once the universe grows larger than $a \sim
V^{-1/6}$, its action rapidly becomes exponentially large and its classical
description becomes possible. What is most important to us here is that the
action at the first (quantum mechanical) part of the evolution of the universe
was $O(1)$, i.e. for any $V(\phi)$ there was no exponential suppression of the
probability of quantum creation of the universe associated with tunneling.

It is clear, however, that at least one of the properties of our solution is
not generic: If one adds any matter contribution to the energy-momentum tensor,
then going back in time will result in the curvature singularity. Usually,
kinetic energy of the scalar field $\phi$ gives the leading contribution in
this limit, since it scales as $a^{-6}$, i.e. much faster than the energy
density of relativistic and nonrelativistic matter. Thus one should take into
account the kinetic energy of the scalar field. The corresponding
Wheeler-DeWitt equation is
\begin{equation}\label{complete}
\left[{\partial^2\over \partial a^2}-{6\over a^{2}}{\partial^2\over \partial
\phi^2}+  12a^4 V\right]\Psi(a,\phi) = 0 \ .
\end{equation}
Assuming, as before, that the potential is flat, $V(\phi) = V$, one can
separate the variables and reduce this equation to the equation for the wave
function of the scale factor $\Psi(a)$:
\begin{equation}\label{5a}
\left[{d^2\over da^2}+12(K a^{-2}+  a^4 V)\right]\Psi(a) = 0 \ .
\end{equation}
Here the constant $K$ has the following meaning: $\rho_{{\rm kin}}={1\over
2}\dot\phi^{2} = {K\over a^{6}}$, i.e. $K$ is the kinetic energy density of the
scalar field at the moment when $a = 1$ in Planck units.

\begin{figure}[h!]
\vskip 0.1cm \centerline{\leavevmode\epsfysize= 5 cm \epsfbox{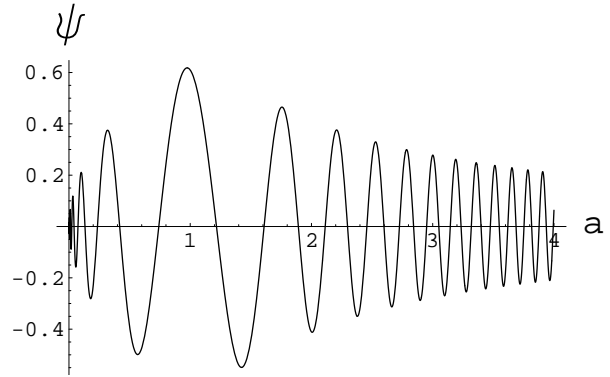}}
\caption[1]{\label{stiffpsi} A typical solution of the Wheeler-DeWitt equation
in the case when the universe initially was dominated by kinetic energy of the
scalar field,  with $V(\phi) = const \ll 1$ and $K \gg 1$. The scale factor is
shown in units of $\left({K\over V}\right)^{{1/6}}$. Inflationary regime occurs
at $a > \left({K\over V}\right)^{{1/6}}$. The wave function $\Psi(a)$ vanishes
at $a \to 0$.}
\end{figure}

At  $a\gg \left({K\over V}\right)^{{1/6}}$ the term $K a^{-2}$ is much smaller
than $a^4 V$, and therefore the solutions of Eqs. (\ref{5a})  look the same as
the solutions of Eq. (\ref{5}), see Fig. \ref{stiffpsi}.  The behavior of the
solutions at  $a\ll  \left({K\over V}\right)^{{1/6}}$  depends on the value of
the coefficient $K$. For $48K < 1$ the solution is non-oscillatory,
\begin{equation}\label{sqrt}
\Psi(a) = \sqrt a\, \Bigl(C_1 \ a^{-{1\over 2}\sqrt{1-48K}} +C_{2}\ a^{{1\over
2}\sqrt{1-48K}}\Bigr)\ .
\end{equation}
The combined solution for all values of $a$ for $48K < 1$ looks very similar to
one of the solutions shown in Fig. \ref{wdw}, approaching $\Psi(a)=0$ at $a\to
0$. Meanwhile, for $48K > 1$ at  $a\ll  \left({K\over V}\right)^{{1/6}}$ one
has an oscillatory solution
\begin{eqnarray}\label{osc}
\Psi(a) &=&\sqrt a\, \Bigl(C_1 \ \sin\left[{1\over 2}{ \sqrt{48K-1}\, \log
a}\right]\nonumber\\ &+&C_{2}\ \cos\left[{1\over 2}{ \sqrt{48K-1}\, \log
a}\right]\Bigr)\ .
\end{eqnarray}

These results can be interpreted in the following way. The condition $48K <1$
means that the kinetic energy can be greater than the Planckian energy only
when the size of the universe is smaller than the Planckian size: $a \lesssim
0.5$. In this case the semiclassical (WKB) approximation works only at $a >
V^{{-1/6}}$. Meanwhile, for $48K\gg 1$ the kinetic energy becomes Planckian
when the universe is large in Planckian units, at $a= K^{1/6} \gg 1$, and the
solution of the WDW equation remains semiclassical and oscillatory  for all
$a$.

The solutions with $K\gtrsim 1$ may be interesting from the point of view of
the probabilistic interpretation of quantum cosmology. One of the main problems
there is the absence of the S-matrix approach. Consider, for example, the
exponential suppression of quantum creation of a closed de Sitter space, Eq.
(\ref{1}). To study tunneling in quantum mechanics one should consider an
incident wave, a reflected wave and a transmitted wave. The total probability
current is conserved, but the current corresponding to the transmitted wave is
exponentially smaller than the current corresponding to the incident wave, the
difference being accounted by the reflected wave.

Meanwhile when one studies creation of a closed universe due to the tunneling
from $a = 0$ to $a = \sqrt {3\over V}$ through the potential barrier ${\cal
V}(a) =6\pi^2 a^2 - 2\pi^2 a^4 V$, one finds that the potential is positive at
small $a$. Therefore the description of tunneling begins under the barrier,
where the wave function does not oscillate, so there are no incident and
reflected waves.  If one adds to ${\cal V}(a)$ the kinetic energy term $-2\pi^2
K/a^2$, the effective potential ${\cal V}(a)$ becomes negative at small $a$,
but one can show that for  $K < 1/192\pi^4$ the solutions of the WDW equation
remain non-oscillatory at small $a$, and the standard description of tunneling
breaks down. A possible interpretation of the exponential suppression (\ref{1})
for the case $K = 0$ can be found in \cite{Vil89}.

However, for $K > 1/192\pi^4$, the WKB approximation is applicable at small
$a$, which leads to the existence of solutions oscillating at $a\lesssim
K^{1/4}$. In terms of the variable $\alpha =\log a$, one can interpret these
solutions as waves coming from $-\infty$ and almost completely reflected back
from the barrier at $a= (K/3)^{1/4}$. In this case one can use the standard
quantum mechanical description of tunneling. Note that for $K \ll V^{-2}$ an
addition of the term $-2\pi^2 K/a^2$ practically does not affect the shape of
the potential barrier at $a \sim H^{-1} \sim V^{-1/2}$, so the final result for
the exponential suppression of tunneling (\ref{1}) remains intact. This allows
us to look at the results of the previous sections from a different point of
view: If a closed universe is born small, with the total kinetic energy of the
scalar field being smaller than $V^{-2}$ at the Planck time, then, at the
classical level, it is going to collapse. It may survive due to tunneling, but
the probability of this event is suppressed by $\exp\Bigl(-{24\pi^2\over
V(\phi)}\Bigr)$.

One can avoid the tunneling if the universe was created with kinetic energy $K
\gtrsim V^{-2}$, but in this case one would need to explain how could it happen
that the universe at the Planck density contained matter with a total mass
greater than $V^{-2}$ Planck masses (which was greater than $10^{16}$ for $V <
10^{-8}$), and why it was homogeneous if it consisted of $V^{-2}$ causally
disconnected domains. As we already discussed in Section \ref{problem}, the
probability of such event for $V \ll 1$ is expected to be exponentially small,
even though under certain conditions it might be greater than the probability
given by the square of the tunneling wave function (\ref{1}); see Eqs. (18) and
(19) in \cite{Linde:1994wt}.

If we repeat the same investigation for the compact flat or open universes, we
will see that for $K \gg 1/48$ the incident waves coming from $\alpha =\log a
=-\infty$ are adiabatically (i.e. practically without any reflection)
transformed into waves moving towards $\alpha =\log a =+\infty$. This result,
which we have verified numerically, is in agreement with our conclusion that
there is no exponential suppression of the probability of creation of such
universes.

\section{Discussion and Summary}

Quantum cosmology is a rather esoteric science. Perhaps for this reason the
authors of one of the pioneering papers on creation of the flat universes with
nontrivial topology \cite{ZelStar} did not want to make any definite statements
concerning the relative probability of creation of a closed universe as
compared to the probability of creation of a compact flat or open universe. In
our opinion, however, the last 20 years of the debates on the probability of
quantum creation a closed universe or an infinite open universe
\cite{Hartle:1983ai,Linde:1983mx,Vilenkin:1984wp,Hawking:1998bn,Linde:1998gs}
demonstrated that in all known cases the probability of creation of a low
energy density inflationary universe is exponentially suppressed. This is a
very important issue since in many inflationary models the energy density of
the universe during inflation is many orders of magnitude smaller than the
Planck density. Meanwhile, as we have seen, there is no exponential suppression
of creation of the low energy density inflationary universe if one considers
compact open or flat universes. It would be interesting to investigate the
behavior of the wave function of the universe for more general anisotropic
cosmological solutions considered in \cite{ZelStar,KLS85}.

From this perspective, one may argue that when the baby universe was born, it
has chosen the path of the least resistance: If inflation may occur only at $V
\ll 1$, it is exponentially more probable that when the universe was born, it
was looking not like a sphere but like a crystal with identified sides. In most
versions of inflationary cosmology, initial anisotropy of the universe was
completely erased by the subsequent long stage of inflation, so it is rather
unlikely that we are going to see any trace of the nontrivial topology of the
universe. However, with a specific fine-tuning of inflationary potential
\cite{Linde:2003hc}, the inflationary stage can be made sufficiently short, in
which case the remnants of the original anisotropy of the universe may become
observable \cite{topol4}.

The possibility that our universe may be a compact open or flat space may have
important implications for string cosmology. String theory is based on the idea
that 6 space dimensions are compactified. Therefore it seems natural to assume
that initially {\it all}\, space dimensions were compact; later some of them
continued expansion, whereas some others become stabilized, e.g. by the KKLT
mechanism \cite{KKLT}. In all known versions of string cosmology, the process
of inflation occurs at a density which is much smaller that the Planck density.
It is quite interesting therefore that the very idea of compactification, being
applied to our universe, may help us to resolve the problem of initial
conditions for inflation in string theory \cite{Blanco-Pillado:2004ns},
including the so-called overshooting problem \cite{brustein,kaloper,cop,brnew}.
It seems very intriguing that creation of the universe may be facilitated by
supersymmetry, which suppresses the Casimir effect. Could it be that
supersymmetry is weakly broken in our universe simply because the probability
of creation of the universes with strongly broken supersymmetry is
exponentially suppressed?

This paper was devoted to the problem of initial conditions for inflation. One
should note, however, that in the eternal inflation scenario
\cite{Vilenkin:xq,Eternal} the problem of initial conditions may become
irrelevant. One may argue that even if the probability of proper initial
conditions for eternal inflation is small, most observers are going to live in
the parts of the universe produced by eternal inflation. Moreover, under
certain assumptions one can show that the probability distribution to live in
the parts with different properties does not depend on the initial conditions
at the moment of the universe creation \cite{LLM}. This is an important
consideration, because inflation in the string theory landscape scenario is
eternal \cite{Susskind,Linde:2004kg}. However, evaluation of probabilities in
eternal inflation scenario is a rather delicate issue, which requires
investigation of measure in quantum cosmology. Moreover, there are many
inflationary models that do not lead to eternal inflation. Therefore it is good
to have an independent argument that in certain cosmological models there is no
exponential suppression of the probability to have inflation with $V\ll 1$.

Independently of the issues related to eternal inflation, our investigation
leads us to a rather unexpected conclusion.  If inflation is eternal, the
universe should look like an eternally growing fractal \cite{Aryal:1987vn,LLM}.
If inflation is not eternal, then our investigation suggests that most probably
we live in a flat or open compact universe with nontrivial topology.  None of
these possibilities correspond to the standard textbook models of a closed
universe or of an infinite flat or open universe.

We are planning to return to some of the problems discussed above in a separate
publication \cite{KLS}.

\

I am grateful to R.~Kallosh, N.~Kaloper,  V.~Mukhanov, and S.~Shenker for
enlightening discussions. I am especially grateful to L.~Kofman, J.~Martin, and
A.~Starobinsky for sharing with me many of their own ideas on the issues
discussed in this paper. This work was supported by NSF grant PHY-0244728.

\end{document}